\documentclass[10pt]{article}
\usepackage{amsmath,amsfonts,amsthm,amssymb,amscd}
\usepackage[toc,page]{appendix}
\numberwithin{equation}{section}

\newcommand{\be}{\begin{equation}}
\newcommand{\ee}{\end{equation}}
\newcommand{\bs}{\begin{split}}
\newcommand{\es}{\end{split}}
\newcommand{\ba}{\begin{align}}
\newcommand{\ea}{\end{align}}
\newcommand{\basl}[1]{\begin{align}\begin{split}\label{#1}}
\newcommand{\bas}{\begin{align}\begin{split}}

\newtheorem{theo}{Theorem}[section]
\newtheorem{prop}[theo]{Proposition}
\newtheorem{lemm}[theo]{Lemma}

\newcommand\R{\mathbb{R}}
\newcommand\C{\mathbb{C}}

\title{
Lindblad approximation and spin relaxation \\in quantum electrodynamics
 }

\author{ L. Amour, J. Nourrigat}

\date{Universit\'e de Reims, France}

\begin{document}

\maketitle

\begin{abstract}
\noindent
This article is concerned with the time evolution of a $\frac{1}{2}$-spin particle in a constant external magnetic field with the quantized electromagnetic field (photons). We derive a Lindblad (or GKLS) type approximation  of the spin dynamics together with a precise control of the error coming from this approximation. The error term is  bounded by $g^2$ where $g$ is the coupling constant of the spin-photon interaction. The point here is the uniformity in  time $t>0$ of this error control.
\end{abstract}

\parindent=0pt

\


\


\parindent=0pt
\parindent = 0 cm

\parskip 5pt
\baselineskip 12pt

 \section{Introduction and statement of the result.}    \label{s1}

\subsection{Introduction.}

Nuclear magnetic resonance (NMR) is the interaction  with a constant magnetic field of one or several $\frac{1}{2}$-spin particles fixed at different points of $\R^3$.
NMR is studied by Bloch \cite{BL} in 1946 including the following two aspects. On the one hand, the spin is viewed as a vector in $\R^3$ with a Larmor precession about the constant field. The spin time evolution follows the so-called Bloch equations. On the other hand, ad hoc additional terms are included in the Bloch equations in order to obtain the spin relaxation, that is, the property of the spin pushed away from its equilibrium position to come back,  in contrast to the initial Bloch equations.  Note that, these additional terms  come from  thermal agitation and interaction with other nuclei.

Our aim in this paper is to show that the spin relaxation can still be explained by a model of NMR in the framework of quantum electrodynamics (QED), even  if these fundamental physical phenomenons (thermal agitation and nuclei interaction) are not included in the model.

Such a model is associated with a Hamiltonian operator introduced by Cohen-Tannoudji, Dupont-Roc and  Grynberg \cite{CTD01} (see also Reuse \cite{Reu}) and is a particular case of the generalized spin boson model. These model and Hamiltonian are called here CTDRG model and Hamiltonian. We were concerned to show that the initial Bloch equations are an approximation of this Hamiltonian and our interest here is to prove that the spin relaxation is another approximation. These two distinct approximations of the same Hamiltonian occur in two different regimes. The first approximation comes from the semiclassical expansion.
The Bloch equations are the semiclassical approximation of the CTDRG Hamiltonian (see \cite{A-J-N-2}).  The second approximation is a form introduced by  Gorini, Kossakowski and Sudarshan \cite{G-K-S} and also by Lindblad \cite{Lind}. According to the terminology of \cite{C-P}, we can call GKLS operator (respectively GKLS approximation) this type of operator (respectively approximation). It is often used, not  for our model but for open quantum systems \cite{Haa73,Dav,SP,A-L}. It is the goal of this work  to show that the spin relaxation is a GKLS approximation of the CTDRG Hamiltonian.

This article is concerned with the case of one fixed $\frac{1}{2}$-spin particle, the case of molecules will be analyzed in a subsequent work. 
In this simplified case, the physical system is constituted, on the one hand, of a spin  fixed at the origin interacting with an external magnetic field  and on the other hand, of the quantized electromagnetic field.  
The model of  CTDRG (see also Reuse) will be detailed in Section 2.   The state Hilbert space of the system is the completed tensor product  ${\cal H }_{ph}  \otimes {\cal H }_{sp}$ where $ {\cal H }_{ph}$ is the space for photons and  $ {\cal H }_{sp}$  is the space for the spin. The Hamiltonian of the system is a selfadjoint  operator  $H(g)$ defined in Section 2, depending on a parameter $g$ being the coupling constant.

We are concerned with the evolution  at time $t>0$ of any given linear operator $\sigma\in  {\cal L} ({\cal H}_{sp})$ (spin observable), namely:
\be\label{S-of-t}  S (t,  \sigma ) =   e^{ i tH(g)}  (I \otimes \sigma  )  e^{ -i tH(g) }.\ee

More precisely, we are interested in the average value of these observables with initial states in the photon vacuum.
The state $\Psi_0 $ stands for the vacuum in  ${\cal H}_{ph}$ which is a Fock space (see Section 2).
One defines an operator $\sigma _0 (T)$ in  ${\cal L} ({\cal H}_{sp})$ with any arbitrary operator  $T$ in ${\cal L} ({\cal H}_{ph} \otimes  {\cal H}_{sp})$
by:
\be\label{sigma-0}  < \sigma _0(T) a, b> \,= \,< T (\Psi_0 \otimes a) , (\Psi_0 \otimes b) >, \ee
for all $a$ and $b$ in ${\cal H}_{sp}$.
In particular, for any spin observable $\sigma \in {\cal L} ({\cal H}_{sp})$, for all $a\in {\cal H}_{sp}$ and for each $t>0$, 
 $< \sigma _0 ( S(t , \sigma) ) a, a>$ stands for the average value of the  spin observable
 $\sigma$  at time $t$, with initial state  $\Psi_0 \otimes a$ being in the photon vacuum.
Thus, it is our objective to study an approximation of this average value, namely $ \sigma _0 ( S(t , \sigma) ) $, as the coupling constant $g$ goes to zero.
Our main purpose here is then  to construct an operator    $L$ in  ${\cal L} ( {\cal L} ( {\cal H}_{sp}))$ 
satisfying for every  unitary $\sigma$ in ${\cal L} ( {\cal H}_{sp})$ and for {\sl all} $t>0$:  
\be\label{approx-Lind}  \Vert \sigma _0 ( S(t , \sigma) ) - e^{  t g^2 L } \gamma _t \sigma \Vert
\leq C g^2, \ee 
where $\Vert\cdot\Vert$ stands for the ${\cal L} ( {\cal H}_{sp})$ norm and where $C>0$ is a positive constant independent of $t>0$.
In (\ref{approx-Lind}), $ \gamma _t : {\cal H}_{sp} \rightarrow {\cal H}_{sp}$ refers to the unitary group 
of the spin free time evolution  without quantized electromagnetic field, that is, the Larmor precession and will be recalled below in  (\ref{gamma-t}).  One also notes that the initial observables in (\ref{S-of-t}) are chosen
as $I \otimes \sigma$ and not as $P_0
 \otimes \sigma$ where $P_0$ denotes the projection on the photon
 vacuum like in some earlier works (see, $e.g.$, \cite{SP}). 

The operator $L$ will be defined in (\ref{def-Lind}) below. One then  recognizes in (\ref{def-Lind})  the general form of GKLS operators (see \cite{Lind,G-K-S} 
and also \cite{C-P,Kos72}).  These type of operators together with the associated semigroups are studied in \cite{G-K-S,Lind, FFFS,A-L}.  GKLS approximations for similar but more specific models such as the two-level (spinless) atom in the dipole approximation or  spin boson model   are  considered in \cite{Haa73,H-S,SP,A-L,RH12} but with a control of the error in the weak coupling limit  (see \cite{VH55,VH57,Dav,Dav76,RH12,SP}) and not in the sense of 
  (\ref{approx-Lind}).
  Also note that, in contrast to the semiclassical approximation or to the weak coupling limit approximation, the estimate in (\ref{approx-Lind}) 
  is uniform in  $t\in(0, \infty)$. 
  
   Experimental data seem to indicate that 
 $\sigma _0 ( S(t , \sigma) )$  should have a limit with an exponential rate of convergence as time goes to infinity. This work proves that   the approximation  $e^{  t g^2 L } \gamma _t \sigma$ has already  itself this latter property, this point coming from the eigenvalues results  for the operator $L$ (see Section 3).   
 We do not know whether  $\sigma _0 ( S (t,  \sigma ))$  has or not a limit 
 as $t$ tends to infinity. In that direction, we point out  the works of  H\"ubner Spohn \cite{H-S} and  De Roeck Kupiainen \cite{DRK} for similar models known as  spin-boson model.  For the model studied in \cite{DRK}, the convergence of
    $\sigma _0 ( S (t,  \sigma ))$ toward its limit occurs with a rate of convergence in ${\cal O}(t^{-\alpha})$ for some  $\alpha >0$. We are also unaware whether or not the result of  \cite{DRK} can be applied to the CTDRG model.

\subsection{Construction and properties of the GKLS operator.} 

The GKLS operator is defined with integrals involving the two free  time evolutions  (photons and spin). Let us first mention some general facts concerning these two free dynamics.

 {\it Photon free evolution.}  The photon free Hamiltonian (without interaction) 
 is an unbounded operator  $H_{ph}$ in the space ${\cal H}_{ph}$, both defined in Section 2 together with the  single photon phase space  ${\cal H}_{\C}$.  
 
 We first focus on a particular standard class of operators in  ${\cal H}_{ph}$ known as Segal fields operators.
 We follow all definitions of  \cite{RSII} regarding Segal fields. In particular, $\Phi_S (V)$ is an unbounded operator of the photon  state space ${\cal H}_{ph}$, for any element $V$   belonging to the  single photon phase space ${\cal H}_{\C}$. Then, 
     the three components of the quantized electromagnetic field at a point $x\in \R^3$ are operators written as  $\Phi_S (B_{jx})$  ($j=1,2,3$) where the  $B_{jx}$ are elements of  ${\cal H}_{\C}$  and are defined by (\ref{7.3}) in Section 2. 
In the sequel, we omit the subscript $x$ from the notations since we  shall only use these fields at the origin and we therefore now write  $B_j $ instead of $B_{j0}$. 
   
The time free evolution of an operator (observable) $A$ in
 ${\cal H}_{ph}$ is defined by $ e^{   i t H_{ph} } A
 e^{ -  i t H_{ph}}  $.   Recall that for  Segal fields, the free evolution has the particular simple form:
\be\label{evol-free}   e^{   i t H_{ph} } \Phi_S(V)
 e^{ -  i t H_{ph}} = \Phi_S (\chi_t V ), \ee
where $ \chi_t$ is a linear map of the phase  space ${\cal H}_{\C}$ 
and is given by (\ref{chi-t}). 
 
The scalar product $ <  B_j  , \chi_t B_m >$ will play a crucial role and in particular its time estimate for all $t>0$ when $j=m$.  
According to (\ref{chi-t}) together with (\ref{7.3})  fixing $x=0$, one notices that:
 \be\label{ortho}  <  B_j  , \chi_t B_m >  =  0,\quad {\rm if} \ \ \ j\not=m \ee
and in the case $j=m$, one verifies that:
\be\label{u(t)}   <  B_j  , \chi_t B_j >  =   \frac {2} {3} \int _{\R^3} \frac {|\chi(|k|)|^2 |k| }
{ (2\pi)^{3}} e^{- i t |k|}  \ dk  = u(t), \ee
where $\chi$ is the  smooth ultraviolet cut-off function in (\ref{7.3}). See \cite{SP} for considerations on the cut-off function for a more general model, namely the Pauli-Fierz Hamiltonian.  This scalar product is also independent of $j$ and is denoted by $u(t)$ in the rest of the paper.
Then, it is easily checked using integrations by part in the radial variable that, there exists 
 $C>0$ independent of $t$ such that:
 \be\label{majo-u(t)} |u(t)| \leq \frac {C} {1+ t^3 },\quad t>0.\ee

{\it Spin free evolution.} The free time dynamics of a spin  observable $\sigma \in {\cal L}({\cal H}_{sp})$,  
without the quantized electromagnetic field but in the presence of an external magnetic field  ${\bf B}_{ext}  = (B^{ext} _1 ,  B^{ext} _2 ,
B^{ext}_3) \not = 0$, is given by:
\be\label{gamma-t}  \gamma _t \sigma = e^{it H_{mag}}  \sigma  e^{-it H_{mag}},  \ee
where $ H_{mag}$ is defined in (\ref{H-mag}) below. 

For our purpose, it will be actually simpler to use the basis of 
 ${\cal L}({\cal H}_{sp})$ constituted with the eigenvectors of $\gamma _t$ (for fixed $t$).  With the external magnetic field 
 ${\bf B}_{ext} = (0, 0, \beta )$ where $\beta >0$,  the basis is written as:  
  $$ I, \quad\sigma (1) = \frac {1} {\sqrt 2} (\sigma _1 + i \sigma _2),\quad \sigma (0)  = \sigma_3,\quad
   \sigma (-1) = \frac {1} {\sqrt 2} (\sigma _1 - i \sigma _2). $$
Note that $\sigma (m)^{\star} = \sigma (-m)$.  One observes that:
   \be\label{gamma-t-sima}  \gamma _t \sigma (m) = e^{2i m \beta t } \sigma (m),\quad
   m = \{ 1, 0, -1 \}. \ee 
We shall often use the notation  $ I = \{ 1, 0, -1 \} $ in sequel. 

   {\it Definition of the GKLS operator.} Set:
   \be\label{d(m)}  d_m = \int _0 ^{\infty } u(t)  e^{2i m \beta t }  dt,\quad  m \in I. \ee
The GKLS operator is defined by:
    \be\label{def-Lind} L(A) = \sum _{m\in I}  ( {\rm Re } \ d_m)  \Big  [ \sigma (m)  A \sigma (m)^{\star}  - \frac {1} { 2}
    [ A , \sigma (m) \sigma (m)^{\star}  ] _+ \Big ] - \frac {i} { 2} [A , H_L ],  \ee
    where
    \be\label{def-Lind-2} H_L =   \sum _{m\in I}  ({\rm Im } \ d_m)    \sigma (m) \sigma (m)^{\star},
     \ee 
for any $A\in {\cal L}({\cal H}_{sp})$ (recall that $\sigma (m)^{\star} = \sigma (-m)$). 
We also use the notation  $[A , B ] _+ = AB + BA$. Then, one recognizes   in (\ref{def-Lind})(\ref{def-Lind-2}) the standard form of GKLS operators.

 Suppose that the external magnetic field is ${\bf B}_{ext} = (0, 0, \beta)$ with $\beta >0$. Then, the only hypothesis assumed in this work is the following one (Fermi Golden Rule):
$$ \chi (2 \beta ) > 0, \leqno (FGR) $$
where the function $\chi$ is the smooth ultraviolet cutoff (see (\ref{7.3})).

Our first result concerns the eigenvalues of the GKLS operator.
 
 \begin{prop}\label{signe}   Suppose that the external magnetic field is  ${\bf B}_{ext} = (0, 0, \beta )$ with $\beta >0$. Then,
the operator  $L$ mapping ${\cal L}({\cal H}_{sp})$  into ${\cal L}({\cal H}_{sp})$
has  $I$, $\sigma (1)$, $\sigma (0) + I $ and $\sigma (-1)$ as eigenvectors. 
One has $L(I)= 0$ and the three other eigenvalues have nonpositive real parts. Under the hypothesis (FGR), these three other eigenvalues have negative real parts 
 
 \end{prop} 

This Proposition is proved in Section 3. We next state to the main result of the paper.

\begin{theo}\label{T-princ}  Assume  that the external magnetic field is  ${\bf B}_{ext} = (0, 0, \beta )$ where $\beta >0$ and 
suppose that the hypothesis (FGR) is satisfied. Then, there exists $C>0$ such that, for all
$\sigma $ in ${\cal L} ({\cal H}_{sp})$ with unit norm and for any $t>0$, the inequality (\ref{approx-Lind}) holds true.

\end{theo}

The proof of this Theorem is completed in Section 4.

    \section{The model.}\label{s2}  

 The Hilbert space of the states of the system under consideration is the completed tensor product
 ${\cal H}_{ph} \otimes {\cal H} _{sp}$ where  ${\cal H} _{ph}$ and ${\cal H} _{sp}$ are respectively the Hilbert spaces of the
 free photons and of the spin particle.

{\it Photons.} The single photon  Hilbert  space ${\cal H}$  is the set of maps
$f \in L^2 (\R^3, \R^3)$ satisfying $k\cdot f(k) = 0$ almost everywhere in $k\in\R^3$
and  where $|f|^2=\int_{\R^3}|f(k)|^2dk$.  One denotes by $<f, g>$ the scalar
product of two elements $f$ and  $g$ of ${\cal H}$. The mapping
$g \rightarrow < f, g>$  is here chosen to be antilinear.  
The photon phase space is ${\cal H}^2$ and is often identified to the complexified space ${\cal H}_{\bf C}$.  
The Hilbert space ${\cal H}_{ph}$ of photon
quantum states is  the symmetrized Fock space
 over  ${\cal H}_{\bf C}$ denoted by ${\cal F}_s ({\cal H}_{\bf C})$.
We also follow \cite{RSII}   for Fock space considerations and notations, in particular for the
usual operators, $\Phi_S(V)$, $\Gamma (T)$ and ${\rm d} \Gamma (T)$,  acting in ${\cal H}_{ph}$, for any $V$ in
 ${\cal H}_{\bf C}$ and any
operator $T$ acting in ${\cal H}_{\bf C}$. The vacuum in ${\cal H}_{ph}$ is here denoted by 
$\Psi_0 $. 

Let  $M_{\omega}$ be the operator with domain   $D(M_{\omega}) \subset {\cal H}$ and defined by  $M_{\omega} q (k) = |k| q(k)$ almost everywhere in $k\in\R^3$.
We denote in the same way the analogous operators defined  on ${\cal H}^2$ or
on the complexified  space ${\cal H}_{\bf C}$. In the Fock space framework, the photon free
energy Hamiltonian operator
$H_{ph}$  is defined as  $  H_{ph} =  {\rm d} \Gamma (M_{\omega})$.

The three components of the magnetic field at a point $x\in \R^3$ are defined
using the elements $ B_{mx}$ belonging to  ${\cal H}_{\bf C}$ and written as:
\be\label{7.3} B_{mx}(k) = {i\chi(|k|)|k|^{1\over 2} \over (2\pi)^{3\over 2}}
e^{-i(   k\cdot x   )} {k\times e_m\over |k|},\quad k\in\R^3\backslash\{0\},\ee
where the function $\chi $ (smooth ultraviolet cutoff) belongs to ${\cal S} (\R)$ and where $(e_m)$ is the canonical basis of $\R^3$.
Then  the operators corresponding to the three components of the  magnetic field 
 at each point  $x$ of $\R^3$ are the $\Phi_{S } (B_{mx}),$ for $m=1,2,3$.

For any $V\in H_{\C}$, equality (\ref{evol-free}) holds with:
\be\label{chi-t} (\chi_t V) (k) = e^{it|k|} V(k),\quad k\in\R^3. \ee

{\it Spin.}  The single $\frac{1}{2}$-spin   space  is  
${\cal H}_{sp} = \C^2   $.
 Let  $\sigma _j$ ($1 \leq j \leq 3$)
be the Pauli matrices:
\be\label{Pauli} \sigma_1 = \begin{pmatrix}  0 & 1 \\ 1 & 0   \end{pmatrix},
\quad
\sigma_2 = \begin{pmatrix} 0 & -i \\ i & 0   \end{pmatrix},
\quad
\sigma_3 = \begin{pmatrix}  1 & 0 \\ 0 & -1  \end{pmatrix}.\ee
The spin Hamiltonian without the quantized field and subject to the external constant magnetic field 
 ${\bf B}_{ext}  = (B^{ext} _1 ,  B^{ext} _2 , 
B^{ext}_3) \not = 0$ is:
\be\label{H-mag}     H_{mag} =    \sum _{m=1}^3
  B^{ext}  _m \  \sigma_m.\ee

{\it The Hamiltonian.}  This Hamiltonian is sometimes  used for modeling NMR in quantum field
theory (see   Section 4.11 of \cite{Reu} and also \cite{Ro-Au,J-H}).  It is a selfadjoint
extension of the following operator initially defined  on a dense subspace of
${\cal H}_{ph} \otimes {\cal H}_{sp} $:
\be\label{7.1} H(g) = H_{ph} \otimes I  + I \otimes  H_{mag}  +  g   H_{int}   ,\ee
where $H_{ph}$ acts in
a domain $D(H_{ph}) \subset {\cal H}_{ph}$,  $g$ is a positive constant
and:
 \be\label{interact}  H_{int}  =  \sum _{m=1}^3
   \Phi_{S } (B_{m0})   \otimes  \sigma_m. \ee

Let us recall the following points concerning domain issues. If an element $U\in {\cal H}_{\bf C}$  belongs to the domain  $ D( M_{\omega }^{-1/2})$
then the Segal  field  $\Phi_S(U)$ is bounded from  $D(H_{ph})$ into ${\cal H}_{ph}$ (see $e.g.$, Proposition 3.4$(ii)$ in
\cite{A-L-N-2}  or  {\cite{DG}}).
This is therefore also valid for the operators $\Phi_{S } (B_{m0})$ in view of
the assumptions on the ultraviolet cutoff function $\chi$ in  (\ref{7.3}).
Thus,   $H(g)$ has a selfadjoint
extension with the same domain as the free  Hamiltonian  $H_0 = H_{ph } \otimes I+ I \otimes  H_{mag}$ domain, according to the Kato-Rellich Theorem.

\section{Nonpositivity of the GKLS operator.}\label{s3}

We first recall the following result.
\begin{lemm}\label{L1}  Let:
\be\label{F(k)}  F  (k) = \frac {2} {3}
 \frac { \chi (|k|) ^2 |k|} {(2 \pi)^3} ,\quad k\in\R^3, \ee 
 where $\chi$ is the ultraviolet cutoff.
If  $\beta >0$ then:
$$ \lim _{\varepsilon \rightarrow 0_+}  \int _{\R^3 \times \R_+ } F(k)
 \cos ( t ( |k| - 2 \beta ) ) e^{- \varepsilon t } dk  dt = \pi \int _
{|k| = 2 \beta } F(k) d\mu (k), $$
where  $\mu $ is measure on the sphere $|k|=2\beta$. If $\lambda \geq 0$ then:
$$ \lim _{\varepsilon \rightarrow 0_+}   \int _{\R^3 \times \R_+ } F(k)
 \cos ( t ( |k| + \lambda  ) ) e^{- \varepsilon t } dk  dt = 0. $$

\end{lemm}

This Lemma directly implies the next Proposition.

\begin{prop} \label{P3.2}The integrals $d_m$ defined in (\ref{d(m)}) satisfy the following identities. 

We have:
$$ {\rm Re} \ d_1 = \pi \int _
{|k| = 2 \beta } F(k) d\mu (k)     \geq 0,  $$ 
where $F$ is the function defined in (\ref{F(k)}). We also have:
$$  {\rm Re}  \  d_0 = {\rm Re}\  d_{-1} = 0. $$
Under the hypothesis {\it (FGR)}, we have in addition:
$${\rm Re} \ d_1 >0.$$

\end{prop}

{\it Proof.}  From   (\ref{u(t)})  and   (\ref{d(m)}), one sees:
$$  {\rm Re} \ d_m = \int _{\R^3 \times \R_+ } F(k)
 \cos ( t ( |k| - 2 m \beta ) ) dk  dt.   $$
Proposition \ref{P3.2} is then deduced from Lemma \ref{L1}. $\Box$

We next turn to the proof of 
Proposition \ref{signe}.

\begin{prop} \label{P3.3}The operator $L$ defined by (\ref{def-Lind})(\ref{def-Lind-2}) satisfies the next equalities:
\begin{align*}
     L(I) &= 0 \\
   L( \sigma (1))   &= \big [ -  ({\rm Re }  \ d_1 ) + i  ({\rm Im } \ d_1 - {\rm Im } \ d_{-1} )
    \big ] \sigma (1) \\
     L( \sigma (0))  & = -2 ({\rm Re }  \ d_1 ) \big ( \sigma (0) + I \big )\\
     L( \sigma (-1))   &= \big [ -  ({\rm Re }  \ d_1 ) - i  ({\rm Im } \ d_1 - {\rm Im } \ d_{-1} )
    \big ] \sigma (-1) .
\end{align*}
\end{prop}

{\it Proof.} From the preceding points, the GKLS operator is written as:
$$ L(A) = ({\rm Re }  \ d_1 ) \Big  [ \sigma (1)  A \sigma (-1)  - \frac {1} { 2}
    [ A , \sigma (1) \sigma (-1)  ] _+ \Big ]  - \frac {i} { 2} \sum _{m\in I}  ({\rm Im } \ d_m )  [A , \sigma (m) \sigma (- m) ].
     $$
One knows that:
    $$ \sigma (1) \sigma (-1) = I + \sigma (0), \quad  \sigma (-1) \sigma (1) =  I - \sigma (0),\quad \sigma (0)  \sigma (\pm 1)  = \pm \sigma (\pm 1),
    $$
    $$
 \sigma (0)^2 = I,\quad  \sigma (\pm 1)^2 = 0.  $$
Proposition \ref{P3.3} then follows. $\Box$

\section{Proof of Theorem \ref{T-princ}.}\label{s4}

Here, we also use the notations of \cite{RSII} for the creation and annihilation operators  $a^{\star} (V)$ and $a(V)$, 
for any  $V$ belonging to the photon phase identified with  ${\cal H}_{\C}$.
One has simple expression analog to (\ref{evol-free}) for the creation and annihilation operators, namely:
 \be\label{chi-t-a-star}    e^{   i t H_{ph} } a(V) e^{  - i t H_{ph}} = a (\chi_t V ),\quad
 e^{   i t H_{ph} } a^{\star} (V) e^{  - i t H_{ph}} = a^{\star} (\chi_t V ).
 \ee
In addition, we shall use the following maps: 
\be\label{A(t)} A (t, V ) =  e^{   i t H(g) } (a(V)  \otimes I)    e^{  - i t H(g) }\ee
 \be\label{Astar(t)} A^{\star}  (t, V ) =  e^{   i t H(g) } (a^{\star} (V)  \otimes I)    e^{  - i t H(g) } \ee
and
 \be\label{Sred(t)} S^{red} (t , \sigma ) = S (t , \gamma_{-t} \sigma), \ee
 for any $V\in {\cal H}_{\C}$ and all $\sigma\in {\cal L} ({\cal H} _{sp})$.

Also, one set:
$$ B(1) =   \frac {1} {\sqrt {2} } ( B_1 + i B _2),\quad B(0) = B_3,\quad
B(-1) =  \frac {1} {\sqrt {2} } ( B_1 - i B _2). $$

In the aim of proving Theorem \ref{T-princ}, we first derive a differential equation from the Heisenberg equation for $ S^{red} (t , \sigma )$ 
(Proposition \ref{deriv-spin}). Next, we get integral equations for 
  $A(t, V)$ and $A^{\star}(t, V)$ 
 also coming from the Heisenberg equation  (Proposition \ref{P42}). 
 Then, we combine these three equations and using a standard approximation  (Markov), we obtain an estimation of the error term (Proposition \ref{markov}). The proof of the error control involves in particular Proposition \ref{norme-comm}.
Finally, in Proposition \ref{markov-lind}, we establish the connection between  
the  Markov approximation and the GKLS operator.

\begin{prop}\label{deriv-spin} One has:
\be\label{expr-S't}  \frac {d} {dt} S^{red}   (t ,   \sigma) =  \frac {ig } {\sqrt 2 }  \sum _{m\in I}
  A^{\star} (t , B (-m) )    S ^{red}  (t, [ \gamma_{t}   \sigma (m) , \sigma ] ) +
  S^{red}  (t, [ \gamma_{t}  \sigma (m) ,   \sigma ] )   A (t , B(m) ). \ee
\end{prop}

  {\it Proof.} Clearly:
\begin{align*}
  \frac {d} {dt} S  (t , \gamma_{-t}  \sigma)   &=
 \frac {d} {dt} e^{it H(g) }    ( I \otimes e^{-it (H_{mag} \otimes I) } )
 (I \otimes   \sigma )  ( I \otimes e^{it (H_{mag} \otimes I) } )  e^{-it H(g) }\\
   &=  ig  e^{   i t H(g) } [ H_{int}, ( I \otimes \gamma_{-t}   \sigma  ) ]
  e^{ -   i t H(g) }.
  \end{align*}
One knows (\cite{RSII}):
   \be\label{a-astar-Phi}  a(V) + a^{\star} (V) = \sqrt {2} \Phi_S (V), \ee
for all $V$ in $H_{\C}$. Consequently,   from  (\ref{interact}) and since the mapping $V \rightarrow a(V)$ is
   anti-$\C$-linear, one has with the above notations:
 \be\label{H-int}  H_{int} =  \frac {1} {\sqrt 2 }   \sum _{m\in I }
 \big [ a(B(m)) + a^{\star} ( B(-m)) \big ]  \otimes \sigma (m).  \ee
Therefore:
$$  [ H_{int}, ( I \otimes  \gamma_{-t}  \sigma  ) ] =   \frac {1} {\sqrt 2 } \sum _{m\in I}
 (a^{\star} (B(-m) ) \otimes I) ( I \otimes [ \sigma (m) ,  \gamma_{-t}  \sigma ])
+  ( I \otimes [ \sigma (m) ,  \gamma_{-t}  \sigma ]) (a(B(m) ) \otimes I)      $$
and thus:
$$  \frac {d} {dt} S  (t , \gamma_{-t}  \sigma)   =  \frac {ig } {\sqrt 2 } \sum _{m\in I}
A^{\star} (t, B(-m) ) S(t, [ \sigma (m) ,  \gamma_{-t}  \sigma ])
+   S(t, [ \sigma (m) ,  \gamma_{-t}  \sigma ]) A(t , B(m)). $$
Then, using (\ref{Sred(t)}), one deduces (\ref{expr-S't}). $\Box$

The next proposition gives the time evolution of the observables 
 $a(V) \otimes I$ 
and  $a^{\star } (V) \otimes I$  relying again  on the Heisenberg equation. 

\begin{prop}\label{P42}  One has: 
\be\label{evol-champ1}  e^{   i t H(g) } (a( \chi_{-t}  V)\otimes I)   e^{  - i t H(g) } =
 a (V)\otimes I -   i \frac {g} {\sqrt 2 }  \sum _{p\in I} \int _0^t
 < B (-p)  , \chi_{-s} V  >  S (s , \sigma (p)  ) ds \ee
and
\be\label{evol-champ2}  A(t, V) =  a (\chi_{t}  V)\otimes I
-   i \frac {g} {\sqrt 2 }  \sum _{p\in I } \int _0^t
< B (-p)  , \chi_{t-s} V  >  S (s , \sigma (p)  ) ds \ee
\be\label{evol-astarV}  A^{\star }(t, V)  =  a^{\star }  (\chi_{t}  V )\otimes I
 +  i \frac {g} {\sqrt 2 } \sum _{p\in I } \int _0 ^{t }  < \chi_{t-s} V  , B (p) >
  S (s , \sigma (p)  ) ds, \ee
for any $V\in H_{\C}$ and time $t>0$.

\end{prop}

  {\it Proof.} For each  $V$ in $H_{\bf C}$, we set:
$$ F (t, V ) =  e^{   i t H(g) } (a( \chi_{-t}  V)\otimes I)   e^{  - i t H(g) }.$$
That is, using (\ref{chi-t-a-star}):
$$ F(t, V ) =  e^{   i t H(g) } (e^{-it H_{ph}} \otimes I) ( a (V)\otimes I)
(e^{it H_{ph}} \otimes I)   e^{ - i t H(g) }.$$
Since the operator $I \otimes H_{mag}$ commutes with  $a (\chi_{-t} V )\otimes I$, we deduce that:
$$ F'(t, V) =  ig   e^{   i t H(g) }  [ H_{int} ,  ( a (\chi_{-t} V )\otimes I)  ] e^{  - i t H(g) }.$$
With (\ref{H-int}) and  using \cite{RSII}, one has:  
$$  [ H_{int} ,  ( a  (W )\otimes I)  ]  = -  \frac {1} {\sqrt 2 }  \sum _{m\in I}
  < B (-m)  , W  > (I \otimes \sigma (m) ),    $$
for every $W$ in $H_{\bf C}$.
Consequently:
\begin{align*}
F'(t, V) &=  - i \frac {g} {\sqrt 2 }  \sum _{m\in I }
 < B(-m)  , \chi_{-t} V  > e^{   i t H(g) } (I \otimes \sigma (m) )e^{  - i t H(g) }\\
  &= -  i \frac {g} {\sqrt 2 }  \sum _{m\in I }
 < B(-m)  , \chi_{-t} V  >  S (t , \sigma (m)  ).
 \end{align*}
Since $F(0, V) =  a (V)\otimes I$, we get (\ref{evol-champ1}).
Replacing $V$ by $\chi_{t}  V$, we obtain (\ref{evol-champ2}). $\Box$

 \begin{prop}\label{norme-comm} If $0 < s < t$ then the commutators $[ A(s , B_j ) , S(t , \sigma) ]$
 and $[ A^{\star} (s , B_j ) , S(t , \sigma) ]$ are bounded and there exists $C>0$ independent of  $t$ such that:
 \be\label{commut}  \Vert [ A(s , B_j ) , S(t , \sigma) ] \Vert + \Vert [ A^{\star} (s , B_j ) , S(t , \sigma) ] \Vert \leq C g, \quad 0 < s <t.  \ee

 \end{prop}

 {\it Proof.} According to (\ref{A(t)}), we have:
 $$  A(t , \chi _{s-t} B_j ) = e^{is H(g)} e^{i(t-s) H(g)} a(\chi _{s-t} B_j )
 e^{i(s-t) H(g)} e^{-is H(g)}. $$
 Applying  (\ref{evol-champ1}) when replacing $t$ by $t-s$, one gets:
 $$  e^{   i (t-s) H(g) } (a( \chi_{s-t}  B_j)\otimes I)   e^{  - i (t-s)  H(g) } =
 a (B_j)\otimes I -   i \frac {g} {\sqrt 2 }  \sum _{m=1}^3 \int _0^{t-s}
 < B_m , \chi_{-\theta } B_j  >  S (\theta  , \sigma _m ) d\theta.  $$
It is already seen that the scalar product $< B_m , \chi_{-\theta } B_j  >$  vanishes if $m\not = j$ and
equals to $u (-\theta)$ if $m=j$.  Consequently:
 $$  e^{   i (t-s) H(g) } (a( \chi_{s-t}  B_j)\otimes I)   e^{  - i (t-s)  H(g) } =
 a (B_j)\otimes I -   i \frac {g} {\sqrt 2 }  \int _0^{t-s}
 u (-\theta) S (\theta  , \sigma _j ) d\theta.  $$
One then deduces that:
 $$ A(t , \chi _{s-t} B_j ) = A(s , B_j ) -   i \frac {g} {\sqrt 2 }
  \int _0^{t-s} u (-\theta)  S (s+ \theta , \sigma _j ) d\theta .   $$
Besides, the operators $S(t , \sigma)$
and $A(t , \chi _{s-t} B_j  )$ commutes in view of (\ref{A(t)}) and (\ref{S-of-t}).

Therefore:
 $$ [ A(s , V ) , S(t , \sigma) ]  =   i \frac {g} {\sqrt 2 }
 \int _0^{t-s}  u (-\theta)
   [ S (s+ u , \sigma _j ),  S(t , \sigma) ]  du.   $$
 The operator in  the above right hand side is bounded and its norm is bounded by 
 $C g$,
 where $C$ is independent  of  $t$ since the function $u$ belongs to $L^1(\R)$. One then obtains (\ref{commut} ). $\Box$

We now combine the foregoing propositions to get a differential equation satisfied by     $S^{red}   (t ,   \sigma)$, in which we make an approximation of type usually called Markov  approximation. In the present case, this approximation only consists in replacing  $s$ by  $t$ at two different places in (\ref{G(t)}). 

 \begin{prop}\label{markov}  One has:
 \be\label{ED-Sred}  \frac {d} {dt} S^{red}   (t ,   \sigma)   =
  G_{mark}(t) + T(t)+  R_4(t),\ee
where:
     \be\label{G1(t)} G_{mark} (t)= \frac {1} {2}  \sum _{m\in I }   \int_0^t  \Big [ -  \overline {u( t - s)}  S^{red} ( t , \gamma_s \sigma (- m)  )
   S ^{red}  (t, [ \gamma_t  \sigma (m) , \sigma ] )  \ee
   $$\ \ \ +  u( t - s)
    S ^{red}  (t, [ \gamma_t  \sigma (m) , \sigma ] )  S^{red} ( t , \gamma _s\sigma (- m)  )
    \Big ] ds  $$
    and
 $$\sigma _0 (T(t)) = 0,$$
 with the estimate:
 \be\label{R4}    \Vert R_4 (t)  \Vert \leq  C g^4. \ee
 \end{prop}

 {\it Proof.}  One applies Proposition \ref{deriv-spin},
  (\ref{evol-champ2}) with  $V= B(m)$ and
 (\ref{evol-astarV}) with $V= B(-m)$. One notices that:
 $$ < B (p)  , \chi_{t-s}  B(m)   > =  \left \{ \begin{matrix}
    0 & {\rm if} & m\not=  p  \\ u(t-s) & {\rm if}  & m=  p
  \end{matrix} \right .. $$
One also uses (\ref{Sred(t)}) showing that
 $ S(s , \sigma (m)  ) = S^{red} (s , \gamma _s \sigma (m) )$. 
One then gets:
$$  \frac {d} {dt} S^{red}   (t ,  \sigma) =  T_1(t) + g^2 G(t) $$
with
     $$ T_1(t) = \frac {ig } {\sqrt 2 }  \sum _{m\in I}
     (a^{\star }  (\chi_{t}  B (-m) )\otimes I)    S^{red} (t, [ \gamma_t \sigma (m) ,  \sigma ] )
     +  S^{red} (t, [\gamma _t \sigma (m) ,  \sigma ] ) (a  (\chi_{t}  B (m) )\otimes I) $$
     and
 \be\label{G(t)}    G(t) = \frac {1} {2}  \sum _{m\in I } \int_0^t  \Big [ -  \overline {u( t - s)}  S^{red} ( s , \gamma _s
  \sigma (- m )
   S ^{red}  (t, [ \gamma_t   \sigma (m) , \sigma ] )  \ee 
   $$+  u( t - s)
    S ^{red}  (t, [ \gamma _t  \sigma (m) , \sigma ] )  S^{red} ( s , \gamma _s \sigma (- m)  )
    \Big ] ds.   $$
  From (\ref{sigma-0}), one has for all operator $A$ and for any $V$ in
    $H_{\bf C}$:
    \be\label{symb-nul} \sigma _0 ( A (a(V)\otimes I ) ) = \sigma _0 ( ( a ^{\star}\otimes I )  (V) A  ) = 0. \ee
Indeed, one has  $a(V) \Psi_0= 0$  since   $\Psi_0$ is the vacuum state (\cite{RSII}). Thus, $\sigma _0  ( T_1(t)) = 0$.  We now approximate  $G(t)$ by the function   $G_{mark} (t)$ defined in (\ref{G1(t)}) (Markov approximation).
   This approximation generates the error:
     $$ G(t) - G_{mark}  (t) = P(t) + P(t)^{  \star } $$
     where:
     $$ P(t) =  \sum _{m\in I }   \int_0^t  u( t - s)
    S ^{red}  (t, [ \gamma _t  \sigma (m) , \sigma ] ) \Big (  S^{red} ( s , \gamma _s \sigma (- m)  )
     - S^{red} ( t , \gamma _s \sigma (- m)  ) \Big ] ds  $$
     $$ = -  \sum _{m\in I }   \int_{0 < s_1 < s_2 < t}   u( t - s_1)
     S ^{red}  (t, [ \gamma _t  \sigma  (m) , \sigma ] )
       \  \partial_{s_2} S^{red} ( s_2  , \gamma _ {s_1}  \sigma (-m)  )  ds_1 ds_2. $$
According to Proposition \ref{deriv-spin} and setting $\tau = \gamma _ {s_1}  \sigma (-m)$, one has:
$$ \partial_{s_2} S^{red} ( s_2  , \tau ) =  \frac {ig } {\sqrt 2 }  \sum _{p\in I } 
  A^{\star} (s_2 , B (p) )    S ^{red}  (s_2, [ \gamma _{s_2}   \sigma (p) , \tau ] ) +
  S^{red}  (s_2, [ \gamma _{s_2}  \sigma (p) ,   \tau ] )   A (s_2  , B (p) ).  $$
In the above product, the term $A^{\star} (s_2 , B_p)$ is not well positioned since it lies between the two spin operators and it would better to have it on the left hand side. One therefore applies 
  (\ref{commut}) replacing $s$ by $s_2$ and  $\sigma $ by
  $[ \gamma _t  \sigma (m) , \sigma ]$. One gets:
  $$ \Vert [ S ^{red}  (t, [ \gamma _t  \sigma (m) , \sigma ] ) , A^{\star} (s_2 , B (p) ) ]
  \Vert \leq C g. $$
One then can write $P(t) = P_1 (t) + P_2(t) $ with:
\begin{align*} P_1(t) &= \frac {ig } {\sqrt 2 }   \sum _{ (m, p) \in I^2 }
  \int_{0 < s_1 < s_2 < t}   u( t - s_1)  A^{\star} (s_2 , B (p) )
  S ^{red}  (t, [ \gamma _t  \sigma (m) , \sigma ] )  S ^{red}  (s_2, [  \gamma _{s_2}  \sigma (p) , \tau ] ) \\&+
  S^{red}  (t , [ \gamma _t  \sigma (m) ,   \sigma  ] )  S ^{red}  (s_2, [ \gamma  _{s_2}  \sigma (p) , \tau ] )  
   A (s_2  , B (p) ) ds_1 ds_2 
   \end{align*}
and
  $$ \Vert P_2(t) \Vert \leq C g^2 \int \int_{0 < s_1 < s_2 < t}   |u( t - s_1)| ds_1 ds_2 \leq C g^2. $$
Let $T_2(t)$ be the operator obtained when replacing $ A (s_2  , B_p)$ by
  $ a( \chi_ {s_2} (B_p)$ and $ A ^{\star} (s_2  , B_p)$ by $ a ^{\star} ( \chi_ {s_2} (B_p)$
in the expression of $P_1(t)$. One has  $\sigma _0 ( T_2 (t))= 0$ from (\ref{symb-nul}).
 Setting $V= B_p$ and replacing
 $t$ by $s_2$  in  (\ref{evol-champ2}) and (\ref{evol-astarV}), one checks:
    $$ \Vert P_1 (t) - T_2(t) \Vert \leq C g^2 \int \int_{0 < s_1 < s_2 < t}   |u( t - s_1)| ds_1 ds_2 \leq C g^2. $$
    The proof of Proposition \ref{markov} is therefore completed with $T(t) = T_1 (t) + T_2 (t) + T_2 (t) ^{\star}$
and with:
    $$ R_4 (t) = g^2 \Big ( P_2(t) + P_1(t) - T_2 (t) \Big ) +
    g^2  \Big ( P_2(t) + P_1(t) - T_2 (t) \Big )^{\star}.$$
    
     $\Box$

    \begin{prop}\label{markov-lind}  The operator $G_{mark} (t)$ defined in (\ref{G1(t)}) 
    and the GKLS operator defined in (\ref{def-Lind})(\ref{def-Lind-2}) satisfy:
      \be\label{R2}   G_{mark} (t) =  S^{red} (t , L\sigma ) + R_2 (t),\quad  \Vert R_2(t) \Vert  \leq \frac {C } {1+ t^2 }. \ee

    \end{prop}

     {\it Proof.} One notes that:
    $$  S ^{red}  (t,  A )  S ^{red}  (t, B) =  S ^{red}  (t, AB),$$
  for any operators $A$ and $B$ in ${\cal H}_{sp}$,
   Thus, $ G_{mark} (t) = S ^{red}  (t, L(t) \sigma )$ with:
    $$ L(t) \sigma  = \frac {1} {2}  \sum _{m\in I }   \int_0^t  \Big [ - \overline {u( t - s)}   \gamma_s \sigma (-m)  
    [ \gamma_t  \sigma (m) , \sigma ]   +  u( t - s)
    [ \gamma_t  \sigma (m) , \sigma ] )   \gamma _s\sigma (- m)  )
    \Big ] ds.   $$
   Using (\ref{gamma-t-sima}), one obtains:  
    $$ L(t) \sigma  = \frac {1} {2}  \sum _{m\in I }   \int_0^t e^{ 2i m \beta s } 
     \Big [ -  \overline {u( s)}  \sigma (-m)  [ \sigma (m) , \sigma ]  +  u(  s) 
      [  \sigma (m) , \sigma ] )   \sigma (- m)  )  \Big ] ds.   $$
    The GKLS operator defined in  (\ref{def-Lind})  is also written as:
    $$ L \sigma  = \frac {1} {2}  \sum _{m\in I }   \int_0^{\infty}  e^{ 2i m \beta s }
     \Big [ -  \overline {u( s)}  \sigma (-m)  [ \sigma (m) , \sigma ]  +  u(  s)
      [  \sigma (m) , \sigma ] )   \sigma (- m)  )  \Big ] ds.  $$
  Taking  (\ref{majo-u(t)}) into account,  one obtains the estimate in (\ref{R2}). \ $\Box$

    {\it End of the proof of Theorem \ref{T-princ}.} In view of the two foregoing Propositions, one can write:
 \be\label{equa-finale}     \frac {d} {dt} S^{red}   (t ,   \sigma)   =
  S^{red} (t , L\sigma ) + R_2 (t) + T(t)+  R_4(t) \ee 
 with the  two estimates (\ref{R2})(\ref{R4}) and with the identity  $\sigma _0 (T(t)) = 0$.
Let $\sigma $
 be one of the eigenvectors of $L$, that is, $I$, $\sigma _1 + i
    \sigma _2$ and $\sigma _3 + I $. If $\sigma = I$ then Theorem \ref{T-princ} is obvious. If $\sigma\not = I$, let   $\mu $ be the  eigenvalue of $L$ associated with the eigenvector $\sigma$. Under the hypothesis {\it(FGR)}, we have ${\rm Re} \mu <0$ 
according to Proposition \ref{signe}. 
Using  (\ref{equa-finale}) and since $\sigma _0 ( T(t)) = 0$, we also have:
   $$  \frac {d} {dt} \sigma _0 ( S^{red}   (t ,   \sigma) )  =
    \mu  g^2  S^{red}   (t ,   \sigma) +
 g^2 \sigma _0 (R_2(t))  +  \sigma _0 (R_4(t). $$
Therefore:
   $$ \sigma _0 ( S^{red}   (t ,   \sigma) ) = e^{  t \mu g^2 } \sigma
   + \int _0^t e ^{  (t-s) \mu g^2 } (  g^2 \sigma _0 (R_2(s))  +  \sigma _0 (R_4(s) ) )
   ds.  $$
Consequently:
   $$ \Vert \sigma _0 ( S^{red}   (t ,   \sigma) ) - e^{  t \mu g^2 } \sigma \Vert
   \leq C g^2 \int _0^t e ^{  (t-s) \mu g^2 }  \frac {  ds } { 1+ s^2 }
   + C g^4  \int _0^t e ^{  (t-s) \mu g^2 }  ds  \leq C g^2. $$
The proof of Theorem \ref{T-princ} is then completed. $\Box$

laurent.amour@univ-reims.fr\newline
LMR  FRE CNRS 2011, Universit\'e de Reims Champagne-Ardenne,
 Moulin de la Housse, BP 1039,
 51687 REIMS Cedex 2, France.

jean.nourrigat@univ-reims.fr\newline
LMR  FRE CNRS 2011, Universit\'e de Reims Champagne-Ardenne,
 Moulin de la Housse, BP 1039,
 51687 REIMS Cedex 2, France.

\end{document}